\documentclass[letterpaper, 10 pt, conference]{ieeeconf}  % Comment this line out if you need a4paper
\IEEEoverridecommandlockouts                              % This command is only needed if 
\usepackage{color}
\usepackage[utf8]{inputenc}
\usepackage{amsmath,amssymb}
\usepackage{mathtools}
\usepackage{graphicx}
\usepackage{float}
\usepackage{wrapfig}
\usepackage{algorithmicx,algpseudocode}
\usepackage{algorithm}
\usepackage{amsfonts,dsfont}
\usepackage{mathrsfs}  
\usepackage{booktabs}
\usepackage{hyperref}
\usepackage[]{moresize}
\usepackage{balance}

\newcommand{\norm}[1]{\lVert #1 \rVert}

\renewcommand{\phi}{\varphi}
\renewcommand{\epsilon}{\varepsilon}

\usepackage{hybridsystem}

\DeclareMathOperator*{\argmin}{argmin}

\newcommand{\fmatrix}[1]
{\begin{bmatrix} f{#1}^{r}
(x #1)\\ 
f{#1}^{s}
(x #1) \end{bmatrix}}
\newcommand{\gmatrix}[1]{\begin{bmatrix} g{#1}^{r}_1
(x {#1}) & g{#1}^{r}_2
(x {#1}) \\
    0   
    & g{#1}^{s}
    (x {#1})   \end{bmatrix}}

\title{\LARGE \bf 
Model-Dependent Prosthesis Control with Interaction Force Estimation}
\author{Rachel Gehlhar and Aaron D. Ames
\thanks{*This material is based upon work supported by the National Science Foundation Graduate Research Fellowship under Grant No. DGE‐1745301, NSF NRI Grant No. 1924526, and CMMI award 1923239.}
\thanks{R. Gehlhar and A. Ames are with the Department of Mechanical and Civil Engineering, California Institute of Technology, Pasadena, CA 91125 USA. Emails:
{\tt\small $\{$rgehlhar, ames$\}$@caltech.edu}}
}

\begin{document}

\maketitle

%%%%%%%%%%%%%%%%%%%%%%%%%%%%%%%%%%%%%%%%%%%%%%%%%%%%%%%%%%%%%%%%%%%%%%%%%%%%%%%%
\begin{abstract}
Current prosthesis control methods are primarily model-independent --- lacking formal guarantees of stability, relying largely on heuristic tuning parameters for good performance, and neglecting use of the natural dynamics of the system. Model-dependence for prosthesis controllers is difficult to achieve due to the unknown human dynamics. We build upon previous work which synthesized provably stable prosthesis walking through the use of rapidly exponentially stabilizing control Lyapunov functions (RES-CLFs). This paper utilizes RES-CLFs together with force estimation to construct model-based optimization-based controllers for the prosthesis. These are experimentally realized on hardware with onboard sensing and computation.
This hardware demonstration has formal guarantees of stability, utilizes the natural dynamics of the system, and achieves superior tracking to other prosthesis trajectory tracking control methods.
\end{abstract}

%%%%%%%%%%%%%%%%%%%%%%%%%%%%%%%%%%%%%%%%%%%%%%%%%%%%%%%%%%%%%%%%%%%%%%%%%%%%%%%%

\section{Introduction}

Powered prostheses are generally controlled by model-independent methods such as impedance control \cite{DesignControlTransProsth, VirtConsCtrlProst, ConfigPowKnee}. These methods rely on heuristic tuning methods to achieve good behavior, lack formal guarantees of stability, and do not utilize the natural dynamics of the system. In \cite{azimi2017robust}, some model-dependence was incorporated into prosthesis control methods to achieve a robust controller. However this method did not account for the interaction force between the human and the prosthesis, which acts as an input to the prosthesis dynamics. The methods in \cite{HybInvStabFeedLin, GreggStableRobust} accounted for the interaction force in constructing a feedback linearizing controller for a prosthesis that was demonstrated in simulation. Generalizing these ideas, in \cite{gehlhar2019control, gehlhar2021separable}, the authors introduced the notion of separable systems and defined a class of RES-CLF controllers to yield provably stable hybrid periodic orbits for separable systems with zero dynamics.

CLFs provide formal guarantees of stability and RES-CLFs \cite{ames2014rapidly} in particular give strong enough conditions for hybrid systems (systems with impacts) with zero dynamics (uncontrollable dynamics) \cite{westervelt2018feedback}. Quadratic programs (QPs) provide a means to implement a CLF constraint while optimizing a cost and provide a flexible framework to incorporate feasiblity constraints such as torque bounds.
CLFs in QPs have been realized in simulation in various works \cite{hereid2014embedding, nguyen2015optimal, kolathaya2016time, gehlhar2021separable}, but few to date on hardware \cite{galloway2015torque}. One difficulty in implementing these controllers on hardware is the typical required inversion of the inertia matrix, which is computationally expensive and prone to error. An alternative CLF-QP was developed in \cite{reher2020inverse} using an inverse dynamics (ID) approach to overcome this challenge and achieved
dynamic crouching behavior in experiment on a 3D underactuated compliant bipedal robot. This ID-CLF-QP is the starting basis for developing an implementable CLF-QP on our robotic prosthesis.

When trying to implement a CLF-QP on a prosthesis, an additional challenge arises since the human dynamics are unknown. While \cite{zhao2014quadratic} applied a CLF-QP to a prosthesis, this was done in a model-independent fashion and required a feed-forward impedance control input term to overcome the limitations of the model-independent nature. 
To implement a model-based prosthesis controller, knowledge of the interaction force between the human and prosthesis is required. While force sensors could provide these measurements, they are expensive, noisy, and not robust to the multi-directional forces and torques present in walking. 
These conditions of force sensors pose implementation issues for using their measurements directly as real-time feedback and restrict prosthesis controllers from being fully model-dependent.

\begin{figure} [t] \label{fig:ampro3} 
\centering
\includegraphics[width=1\columnwidth]{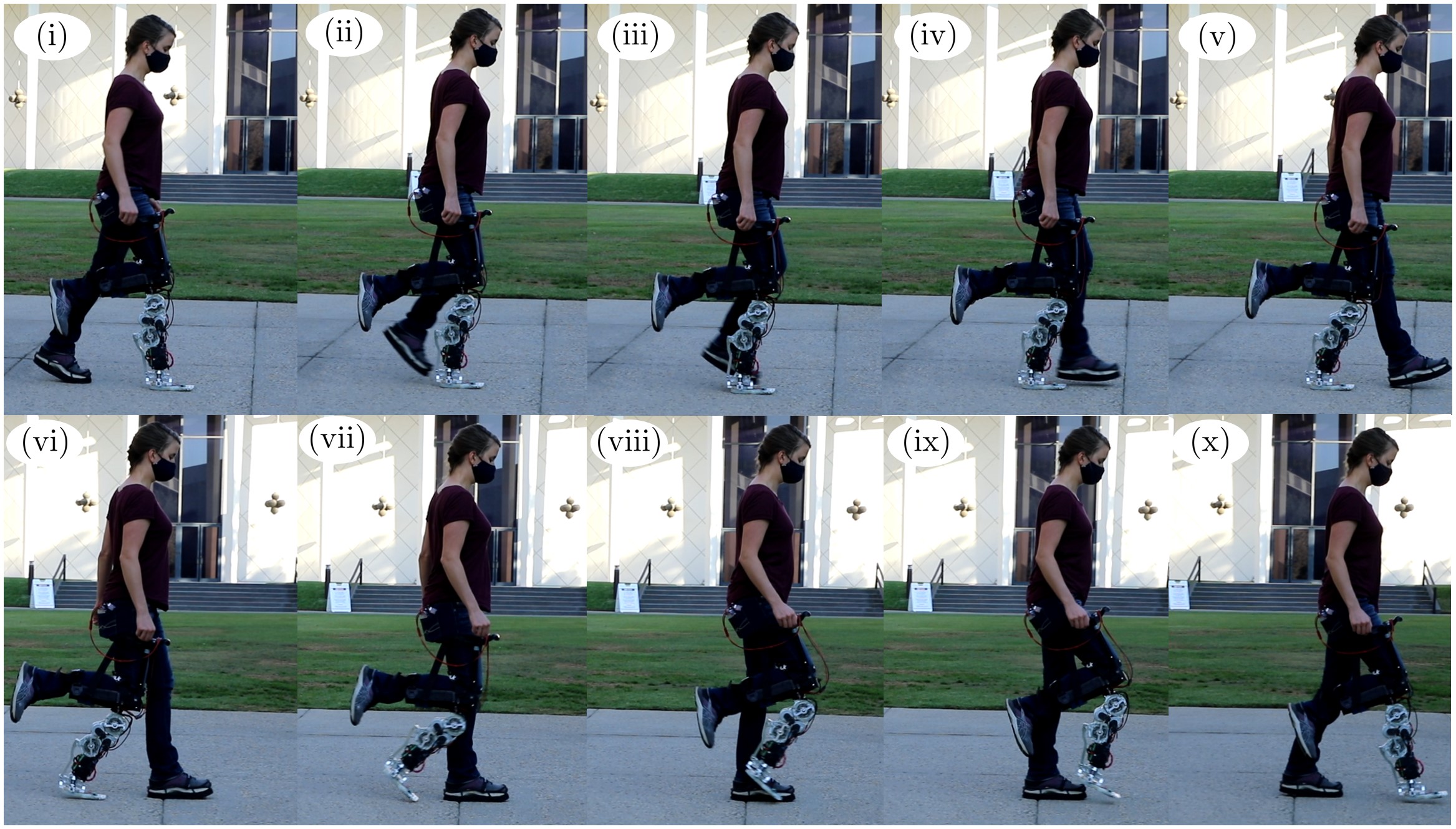}
\vspace{-0.8cm}
{\caption{Gait tiles of powered prosthesis AMPRO3 worn by able-bodied human user walking with model-dependent prosthesis controller. Top shows prosthesis stance, bottom shows prosthesis non-stance. Numbers align with phases of gait trajectory shown in Fig. 5.}}
\vspace{-0.95cm}
\end{figure}

The main result of this paper is the synthesis of model-dependent controllers using force estimation. We leverage RES-CLFs and their formal guarantees in the context of the ID-CLF-QP framework. The unknown dynamics of the human enter the prosthesis dynamics via interaction forces, so we estimate these forces to complete the model-dependent nature. Inspired by the average acceleration discrete algorithm in \cite{AvgAccelDiscrete}, we developed a force estimation method with on-board velocity measurements. To demonstrate these results, we realize the controller on-board the AMPRO3 prosthesis \cite{zhao2017preliminary}, shown in Fig. 1. In particular we demonstrate that the model-based ID-CLF-QP results in accurate tracking. More generally, we are thus able to transfer the formal guarantees afforded by RES-CLFs to hardware, with the result being stable prosthesis locomotion in practice.

In this work, Section \ref{sec:Background} overviews the separable system framework used to develop a RES-CLF for the prosthesis based solely on local information. Section \ref{sec:Control} lays out our specific controller construction for a robotic subsystem. This section describes the discrete force estimation method and how the controller is respectively formed to be incorporate this estimate. The force estimate completes the prosthesis subsystem dynamics to enable model-dependent prosthesis control which we demonstrate in simulation in \ref{sec:Simulation} and experiment in \ref{sec:Experimentation}, yielding provably stable human-prosthesis walking. The main contribution of this paper is the first realization of fully model-dependent prosthesis control, bringing the human into the loop of prosthesis control with strong formal guarantees of stability. 

\section{Background on Separable Subsystems} \label{sec:Background}
To support the main contribution of this paper, 
we first review how these formal guarantees are established for a robotic subsystem like the prosthesis. Since the prosthesis is connected to a human, it is not an independent system, but rather a subsystem of a larger system, of which it does not have full information. We show how to develop a class of controllers for a robotic subsystem that use only local information but lead to stability guarantees for the full-order separable robotic system. 

\newsec{Robotic Control System.}
For an $\eta$ DOF robotic system in 2D space, consider the coordinates $q = (q_l^T, q_f^T, q_s^T)^T \in \mathbb{R}^\eta$ which define the configuration space $\mathcal{Q}$. To create a separable robotic system, 
we consider the portion of the system defined with coordinates $q_s \in \mathbb{R}^{\eta_s}$ to be a robotic \textit{subsystem} that is rigidly attached with a 3 DOF fixed joint ($x, z$ Cartesian position, and pitch), with coordinates $q_f$, to the rest of the system with coordinates $q_l \in \mathbb{R}^{\eta_l}$, where $\eta_l + \eta_s + 3 = \eta$. The subsystem has $m_s$ actuators and the rest of the system has $m_r$. This Euler-Lagrange equation with positional constraints gives the dynamics \cite{MLS}:
\begin{align} \label{eq:robotDynamics}
    & D(q) \ddot{q} + H(q, \dot{q}) = Bu + J_h^T(q) \lambda_h 
    \\ \label{eq:holoConstr}
    & J_h(q) \ddot{q} + \dot{J}_h(q, \dot{q})\dot{q} = 0.
\end{align}
Here $D(q)$ is the inertia matrix; $H(q, \dot{q})$ the vector sum of Coriolis, centrifugal, and gravity forces; $B$ the actuation matrix, $\lambda_h = (F_f^T, \lambda_g^T)^T \in \mathbb{R}^{3 + \eta_g}$ the constraint wrenches to enforce the fixed joint and the $\eta_h$ contact holonomic constraints, respectively; and
$J_h(q)$ the Jacobian of the holonomic constraints of the fixed joint and contacts.
These dynamics and constraints can be used to solve for $\lambda_h$ by,
\begin{equation} \label{eq:force}
    \lambda_h = (J_h D^{-1} J_h^T)^{-1} (J_h D^{-1}(H - Bu) - \dot{J}_h \dot{q}).
\end{equation}

\newsec{Robotic Subsystem.}
By defining floating base coordinates $\bar{q}_B \in \mathbb{R}^3$ for the subsystem at the connection point with the rest of the system, we can define the robotic subsystem with its own configuration coordinates $\bar{q} = (\bar{q}_B^T, q_s^T)^T \in \mathbb{R}^{\bar{\eta}}$, with $\bar{\eta} = 3+\eta_s$, and write the constrained subsystem dynamics,
\begin{align} \label{eq:robotSubsystem}
    & \bar{D}(\bar{q})\ddot{\bar{q}} + \bar{H}(\bar{q}, \dot{\bar{q}}) 
    = \bar{B}u_s + \bar{J}^T_h(\bar{q}) \bar{\lambda}_h + \bar{J}^T_f(\bar{q}) F_f
    \\ \label{eq:holoConstrSub}
    &\bar{J}_h \dot{\bar{q}} + \dot{\bar{J}}_h(\bar{q}, \dot{\bar{q}}) \dot{\bar{q}} = 0 
\end{align}
Here $\bar{J}_h(\bar{q})$ is the Jacobian of the $\bar{\eta}_h$ holonomic constraints for the contacts of the subsystem with constraint wrench $\bar{\lambda}_h$, and $F_f$ is the interaction forces and moment (we call interaction force for simplicity) between the subsystems given as input to these subsystem dynamics, projected to the base coordinates with $\bar{J}_f$.

\newsec{Separable Subsystems.}
We can write the robotic full-order dynamics \eqref{eq:robotDynamics} as an ODE using the states $x_q = (q^T, \dot{q}^T)^T$:
\begin{equation*}
    \dot{x}_q = 
    \underbrace{
    \begin{bmatrix}
    \dot{q} \\
    D^{-1}(q)(-H(q, \dot{q}) + J_h(q)^T \lambda_h
    \end{bmatrix}
    }_{f_q(x_q)}
    +
    \underbrace{
    \begin{bmatrix}
    0 \\
    D^{-1}(q)(B)
    \end{bmatrix}
    }
    _{g_q(x_q)}
    u
\end{equation*}
By selecting a different set of states $x = (x_r^T, x_s^T)^T$ with $x_r = (q_l^T, q_f^T, \dot{q}_l^T, \dot{q}_f^T)^T$ and $x_s = (q_s^T, \dot{q}_s^T)^T$, our ODE takes the following form:
\begin{align} \label{eq:sepSyst}
    \begin{bmatrix}
        \dot{x}_r \\ \dot{x}_s
    \end{bmatrix}
    &=
    \underbrace{\fmatrix{}}_{f(x)}
    + \underbrace{\gmatrix{}}_{g(x)}
    \begin{bmatrix}
        u_r \\ u_s
    \end{bmatrix},
    \\ \notag
        x_r \in &\mathbb{R}^{n_r}, \,\,
        x_s \in \mathbb{R}^{n_s}, \,\,
        u_r \in \mathbb{R}^{m_r}, \,\,
        u_s \in \mathbb{R}^{m_s},
\end{align}
The 0 appears in the actuation matrix $g(x)$ because the fixed joint present in this system completely decouples the subsystem dynamics from the actuation $u_r$ of the rest of the system since all the interaction goes through the constraint wrench for the fixed joint. See \cite{gehlhar2019control} for details. Since $q_f \in x_r$, the control input $u_s$ still affects the dynamics of $x_r(t)$. 
In \cite{gehlhar2019control}, this system was defined as a \textit{separable control system}, which has the unique attribute that the dynamics of $x_s(t)$ only depends on $u_s$ and not $u_r$.

We separate this separable system into a \textit{separable subsystem} and \textit{remaining system} \cite{gehlhar2019control, gehlhar2021separable}, defined respectively:
\begin{align} \label{eq:Subsystem}
    \dot{x}_s &= f^s(x, z) + g^s(x, z) u_s, 
    \\ \label{eq:remSys}
    \dot{x}_r &= f^r(x, z) + g^r_1(x, z) u_r + g^r_2(x, z) u_s.
\end{align}

\newsec{Equivalent Subsystem}
We can write the robotic subsystem dynamics \eqref{eq:robotSubsystem} as an ODE following a method similar to that used for the full-order dynamics, but this time only the dynamics of $x_s(t) = (\bar{q}_s^T, \dot{\bar{q}}_s^T)^T$ are used such that we obtain an alternative expression for the dynamics of $x_s(t)$ \cite{gehlhar2019control}:
\begin{align} \label{eq:Subsystem'}
        &\dot{\bar{x}}_s = \bar{f}^s(\mathcal{X}) + \bar{g}^s(\mathcal{X}) u_s,
        \\ \notag
        & \mathcal{X} = (\bar{x}_r^T,\, x_s^T,\, \zeta^T)^T \in \mathbb{R}^{\bar{n}},
\end{align}
Here $\bar{x}_s = x_s$, $\bar{x}_r= (\bar{q}_B^T, \dot{\bar{q}}_B^T)^T \in \mathbb{R}^{\bar{n}_r}$ are \textit{measurable states}, $\mathcal{X}$ is the state vector $\bar{x} = (\bar{x}_r^T, x_s^T)^T$ augmented with the \textit{measurable input} $\zeta = F_f \in \mathbb{R}^{n_f}$.
For this subsystem to equate to \eqref{eq:Subsystem}, there must exist a transformation $T(x) = \mathcal{X}$ that yields the following conditions: $f^s(x) = \bar{f}^s(\mathcal{X})$ and $g^s(x) = \bar{g}^s(\mathcal{X})$ for all $x$. This transformation exists for this robotic system and is given in \cite{gehlhar2019control}. While the separable subsystem \eqref{eq:Subsystem} still depends on the full-order states $x$, this \textit{equivalent subsystem} \cite{gehlhar2019control} only depends on local states and measurable states and inputs. In practice $\bar{x}_r$ can be measured with an IMU and $\zeta$ with a force sensor.

\newsec{Separable Subsystem RES-CLF.}
Now that the subsystem is defined in local coordinates $\mathcal{X}$, a whole class of model-dependent controllers can be constructed for the subsystem in terms of locally available information. In \cite{gehlhar2021separable} we defined a RES-CLF $\bar{V}_\varepsilon^s(x_s)$ for an equivalent separable subsystem such that for all $0 < \varepsilon < 1$ and $\mathcal{X} \in \mathbb{R}^{n_s + \bar{n}_r + n_f}$,
\begin{align} \label{eq:equivRESCLF}
        &\bar{c}_1^s \norm{x_s}^2 
        \leq 
        \bar{V}^s_{\varepsilon}(x_s) 
        \leq \frac{\bar{c}_2^s}{\varepsilon^2} \norm{x_s}^2
        \\ \notag
        &\inf_{u_s \in \mathbb{R}^{m_s}} 
        [\dot{\bar{V}}^s_{\varepsilon}(\mathcal{X}, u_s)] 
        \leq 
        -\frac{\bar{c}_3^s}{\varepsilon} \bar{V}^s_{\varepsilon}(x_s),
\end{align}
where $\bar{c}_1^s,\, \bar{c}_2^s,\, \text{and } \bar{c}_3^s$ are positive constants.
This leads to a class of controllers that satisfy $\dot{\bar{V}}_\varepsilon^s (\mathcal{X}, u_s) \leq -\frac{\bar{c}_3^s}{\varepsilon} \bar{V}_\varepsilon^s(x_s)$:
\begin{equation} \label{eq:subK'}
    \bar{K}_\varepsilon^s(\mathcal{X}) = \{ u_s \in \mathbb{R}^{m_s}: 
    \dot{\bar{V}}_\varepsilon^s (\mathcal{X}, u_s)
    \leq
    -\frac{\bar{c}_3^s}{\varepsilon} \bar{V}^s_{\varepsilon}(x_s) \}.
\end{equation}
The $\varepsilon$ term was included in the formulation of CLFs in \cite{ames2014rapidly}
to give a faster rate of convergence for hybrid systems such that the system and its zero dynamics would not be destabilized by the impacts present in the hybrid system. 

\newsec{Main Theoretic Idea.}
The work of \cite{gehlhar2021separable} proved when a RES-CLF stabilizes the remaining system \eqref{eq:remSys}, any controller $u_s$ in this class $\bar{K}_\varepsilon^s(\mathcal{X})$ guarantees stability for the full-order hybrid system with zero dynamics. Hence, this novel methodology of separating a robotic system and creating an equivalent subsystem enables construction of model-dependent subsystem controllers with only local information while guaranteeing stability of the full-order system and utilizing the natural dynamics. 
For the human-prosthesis system, we assume the human stabilizes itself since central pattern generator research suggests biological walkers exhibit stable rhythmic behavior \cite{Taga1995}, (i.e. have limit cycles), and our class of RES-CLF controllers in \cite{gehlhar2021separable} for the remaining human system includes all stabilizing controllers for these hybrid limit cycles.

\section{Control Methods} \label{sec:Control}
To construct the ID-CLF-QP of \cite{reher2020inverse} we construct a RES-CLF for the robotic subsystem and formulate it in a QP without inverting the inertia matrix. Our ID-CLF-QP has an additional $\bar{J}_f^T(\bar{q}) F_f$ term in the dynamics as in \eqref{eq:robotSubsystem} to account for the interaction force between the subsystems. We finally formulate this controller in a hardware implementable way with force estimation to arrive at the form used to achieve provably stable prosthesis control in experiment.

\subsection{Controller Formulation} \label{ssec:ControllerFormulation}
To construct the ID-CLF-QP of \cite{reher2020inverse}, we form subsystem outputs with which we construct our RES-CLF using the methods of \cite{ames2014rapidly}. We show how the ID-CLF-QP incorporates this RES-CLF without inverting the dynamics.

\newsec{CLF Construction.}
To enforce desired trajectories on our robotic subsystem, we define linearly independent outputs, 
\begin{equation} \label{eq:subOutputs}
    y_s(x_s) = y^a_s(x_s) - y_s^d(\tau(x_s), \alpha)
\end{equation}
where $y_s^a(x_s)$ are the actual outputs and $y_s^d(\tau(x_s), \alpha)$ are the desired outputs defined by parameters $\alpha$ and modulated by the state-based phase variable $\tau(x_s)$ \cite{westervelt2018feedback}. 
For our robotic application, we consider position modulating outputs.
We take the derivatives along $\bar{f}^s(\mathcal{X})$ and $\bar{g}^s(\mathcal{X})$ to relate the outputs to the control input $u_s$:
\begin{equation*}
    \ddot{y}_s= L_{\bar{f}^s}^{2}y_s(\mathcal{X}) + L_{\bar{g}^s}L_{\bar{f}^s} y_s(\mathcal{X}) u_s.
\end{equation*}
Here $L_{\bar{f}^s}^{2}y_s(\mathcal{X})$ and $L_{\bar{g}^s}L_{\bar{f}^s} y_s(\mathcal{X})$ are Lie derivatives \cite{IsidoriNonlinSyst} and $L_{\bar{g}^s}L_{\bar{f}^s} y_s(\mathcal{X})$ is invertible since the outputs are linearly independent. Hence our system is feedback linearizable \cite{IsidoriNonlinSyst} and our feedback linearizing controller is,
\begin{equation} \label{eq:feedlin}
    u_s(\mathcal{X}) = \big(L_{\bar{g}^s}L_{\bar{f}^s} y_s(\mathcal{X}) \big)^{-1} \big( -L_{\bar{f}^s}^{2}y_s(\mathcal{X}) + \nu \big), 
\end{equation}
where $\nu$ is our auxiliary control input and by construction $\nu = \ddot{y}_s$. By applying this control law, our output dynamics are linearized and can be written as a linear system with coordinates $\xi = (y_s^T, \dot{y}_s^T)^T$,
\begin{equation*}
    \dot{\xi} = 
    \underbrace{
    \begin{bmatrix}
    0 & I \\
    0 & 0
    \end{bmatrix}
    }_F
    \xi
    +
    \underbrace{
    \begin{bmatrix}
    0 \\
    I
    \end{bmatrix}
    }_G
    \nu
\end{equation*}
Using this linear system we construct a CLF by solving the continuous time algebraic Riccati equation (CARE),
\begin{equation*}
    F^T P + PF - PGG^T P + Q = 0,
\end{equation*}
for $P = P^T >0$, with the user selected weighting matrix $Q = Q^T >0$. From the method of \cite{ames2014rapidly}, we construct a RES-CLF for our subsystem by the following:
\begin{equation*} \label{eq:subRES-CLF}
    \bar{V}^s_\varepsilon(\xi) = \xi^T 
    \begin{bmatrix}
    \frac{1}{\varepsilon} I & 0 \\
    0 & I
    \end{bmatrix}
    P
    \begin{bmatrix}
    \frac{1}{\varepsilon} I & 0 \\
    0 & I
    \end{bmatrix}
    \xi
    =: \xi^T P^\varepsilon \xi,
\end{equation*}
To obtain our convergence constraint, we take the derivative,
\begin{equation*}
    \dot{\bar{V}}^s_\varepsilon(\xi, \nu) = L_F\bar{V}^s_\varepsilon(\xi) + L_G\bar{V}^s_\varepsilon(\xi) \nu \leq - \frac{1}{\varepsilon} \underbrace{\frac{\lambda_{min} (Q)}{\lambda_{max} (P)}}_{\gamma} \bar{V}^s_\varepsilon(\xi),
\end{equation*}
with Lie derivatives along the linearized output dynamics as,
\begin{align*}
L_F \bar{V}^s_\varepsilon(\xi) &= \xi^T (F^T P_\varepsilon + P_\varepsilon F) \xi,
\\ 
L_G \bar{V}^s_\varepsilon(\xi) &= 2 \xi^T P_\varepsilon G.
\end{align*}

\newsec{ID-CLF-QP+$F_f$.}
To formulate the ID-CLF-QP in terms of $\mathcal{X}$, we write this RES-CLF and its derivative in terms of $x_s$ and $\mathcal{X}$ since $\xi$ depends on $x_s$, through the outputs $y_s(x_s)$ and $\dot{y}_s(x_s)$, and $\nu$ depends on $\mathcal{X}$ through the relationship, obtained from \eqref{eq:feedlin}:
\begin{equation} \label{eq:nu}
    \nu =  L_{\bar{f}^s}^{2}y_s(\mathcal{X}) + L_{\bar{g}^s}L_{\bar{f}^s} y_s(\mathcal{X})    u_s(\mathcal{X}).
\end{equation}
This gives us the subsystem RES-CLF \eqref{eq:equivRESCLF} where $\bar{c}_1^s = \lambda_{min}(P)$. $\bar{c}_2^s = \lambda_{max}(P)$, and $\bar{c}^s_3 = \gamma$.

The expression \eqref{eq:nu} requires multiple inversions of the inertia matrix $\bar{D}(\bar{q})$ and holonomic constraint term $\bar{J}_h(\bar{q})$ which is computationally expensive and prone to numerical error.
To avoid this, we recall $\nu = \ddot{y}_s$ and rewrite the outputs $y_s(x_s)$ in terms of the robotic subsystem's configuration coordinates $\bar{q}$, since it is a positional constraint: $y_s(\bar{q})$.
Shown in \cite{reher2020inverse}, for position-modulating outputs $y_s(\bar{q})$, we can equivalently write $\ddot{y}_s$ as, 
\begin{equation} \label{eq:yddot}
    \ddot{y}_s = \underbrace{\frac{\partial}{\partial \bar{q}} \bigg( \frac{\partial \bar{y}_s}{\partial \bar{q}} \dot{\bar{q}} \bigg)}_{\dot{J}_y(\bar{q}, \dot{\bar{q}})} \dot{\bar{q}} 
    + 
    \underbrace{ \frac{\partial y_s}{\partial \bar{q}}}_{J_y(\bar{q})} \ddot{\bar{q}}
\end{equation}

To find a control input $u_s$ close to the feedback linearizing controller \eqref{eq:feedlin} with PD gains on our output accelerations, $\nu = K_p y^s(x_s) + K_d \dot{y}^s(x_s) := \nu_{\rm{pd}}$, we minimize the difference between \eqref{eq:yddot} and $\nu_{\rm{pd}}$ in our QP cost. 
We also include the holonomic constraints in the cost as soft constraints since these are difficult to satisfy exactly on hardware. Considering the variables $\Upsilon = [\ddot{\bar{q}}^T, u_s^T, \bar{\lambda}_h^T, \delta]^T \in \mathbb{R}^{\eta_v}$, with $\eta_v = \bar{\eta} + m_s + \bar{\eta}_h + 1$, and using the terms, 
\begin{align*}
    &J_c(\bar{q}) = 
    \begin{bmatrix}
    J_y(\bar{q}) \\
    \bar{J}_h(\bar{q})
    \end{bmatrix}
    &\dot{J}_c(\bar{q}, \dot{\bar{q}}) =
    \begin{bmatrix}
    \dot{J}_y(\bar{q}, \dot{\bar{q}}) \\
    \dot{\bar{J}}_h(\bar{q}, \dot{\bar{q}})
    \end{bmatrix},
\end{align*}
we formulate our \textbf{ID-CLF-QP+$F_f$}: 
\begin{equation} \label{eq:ID-CLF-QP+Ff}
\resizebox{0.89\hsize}{!}{$
\begin{aligned}
\Upsilon^\star 
= \mathop {\argmin }\limits_{{\Upsilon \in \mathbb{R}^{\eta_v}}} \,
& \Big|\Big| \dot{J}_c(\bar{q}, \dot{\bar{q}}) \dot{\bar{q}} + J_c(\bar{q}) \ddot{\bar{q}} - \mu^{\rm{pd}} \Big|\Big|^2 
+ \sigma W(\Upsilon) 
+ \rho \delta
 \\ 
\textrm{s.t.} \,\, 
& \bar{D}(\bar{q})\ddot{\bar{q}} + \bar{H}(\bar{q}, \dot{\bar{q}}) 
= \bar{B}u_s + \bar{J}^T_h(\bar{q}) \bar{\lambda}_h + \bar{J}^T_f(\bar{q}) F_f
\\
& L_F\bar{V}^s_\varepsilon(\mathcal{X}) + L_G\bar{V}^s_\varepsilon(\mathcal{X}) \big(\dot{J}_y \dot{\bar{q}} + J_y \ddot{\bar{q}}\big) \leq - \frac{\gamma}{\varepsilon} \bar{V}^s_\varepsilon(\mathcal{X}) + \delta
\\
& -u_{\rm{max}} \leq u_s \leq u_{\rm{max}} 
\end{aligned}
$}
\end{equation}
where $\mu^{\rm{pd}} = (\nu_{\rm{pd}}^T, 0^T )^T$, $W(\Upsilon)$ is a regularization term to make the problem well posed, $\sigma$ and $\rho$ are weighting terms, and $\delta$ is a relaxation term such that the torque bounds $(-u_{\rm{max}}, u_{\rm{max}})$ can always be met. (The arguments on $J_y, \dot{J}_y$ are left out for notational simplicity.) This controller selects the joint accelerations $\ddot{\bar{q}}$, control input $u_s$, and holonomic constraint wrench $\bar{\lambda}_h$ to satisfy the robotic subsystem dynamics \eqref{eq:robotSubsystem} and the subsystem RES-CLF \eqref{eq:equivRESCLF} while optimally aiming to satisfy the holonomic constraints \eqref{eq:holoConstrSub} and smoothly track the desired trajectories. 

\subsection{Controller Realization for Hardware} \label{ssec:ControllerRealization}
Implementing this controller on hardware requires knowledge of the interaction force $F_f$. 
Since a force sensor is not available on the prosthesis platform we developed a method to estimate the interaction force using discrete calculations of acceleration.
We include this estimated term in the dynamics of our QP and realize the QP at sample time to implement on hardware.

\newsec{Force Estimation.}
We estimate  the joint acceleration $\ddot{\bar{q}}^{\rm{est}}$ based on the discrete velocity measurements and time,
\begin{equation*}
    \ddot{\bar{q}}^{\rm{est}}_{k-1} = \frac{\dot{\bar{q}}_k - \dot{\bar{q}}_{k-1}}{t_k - t_{k-1}},
\end{equation*}
where $k$ represents the current time step and $k-1$ represents the previous time step. Finding the difference between our estimated acceleration and the expected acceleration based on the dynamics from the previous time step,
\begin{equation}  \label{eq:qexp}
\resizebox{0.89\hsize}{!}{$
\ddot{\bar{q}}^{\rm{exp}}_{k-1} = \bar{D}(\bar{q}_{k-1})^{-1} \big( - \bar{H}(\bar{q}_{k-1}, \dot{\bar{q}}_{k-1}) + \bar{B} u_{s,k-1} + \bar{J}_h^T(\bar{q}_{k-1}) \bar{\lambda}_{h, k-1}\big),
$}
\end{equation}
we multiply this by the inertia matrix of the previous time step to obtain what we consider the \textit{residual dynamics} $\mathcal{F}_{k-1}$:
\begin{equation}\label{eq:Fk-1}
    \mathcal{F}_{k-1} = \bar{D}(\bar{q}_{k-1})(\ddot{\bar{q}}^{\rm{est}}_{k-1}  - \ddot{\bar{q}}^{\rm{exp}}_{k-1}).
\end{equation}
We essentially back-calculate the interaction force that caused the acceleration difference. 
Note \eqref{eq:Fk-1} cancels $\bar{D}(\bar{q}_{k-1})$ in \eqref{eq:qexp}, such that inertia matrix inversion is not required.
To obtain a smoother signal, 
we average the residual dynamics measurements for $N$ time steps: 
\begin{equation} \label{eq:forceavg}
    \mathcal{F}_{k-1}^{\rm{avg}} = \frac{1}{N} \sum_{i = 1}^N \mathcal{F}_{k-i}.
\end{equation}
By calculating the force projected into joint space, we are smoothing the exact signal we input to the dynamics and do not need a pseudo-inverse of $\bar{J}_f$.

\newsec{ID-CLF-QP+$\mathcal{F}^{\rm{est}}$.} We replace $\bar{J}_f^T F_f$ of \eqref{eq:ID-CLF-QP+Ff} with $\mathcal{F}^{\rm{avg}}_{k-1}$ and evaluate the QP at sample time:
\begin{equation} \label{eq:ID-CLF-QP+Fest}
\resizebox{0.89\hsize}{!}{$
\begin{aligned}
\Upsilon^\star_k = \mathop {\argmin }\limits_{{\Upsilon_k \in \mathbb{R}^{\eta_v}}} \,
&\Big|\Big| \dot{J}_c(\bar{q}, \dot{\bar{q}}_k) \dot{\bar{q}}_k 
+ J_c(\bar{q}) \ddot{\bar{q}}_k - \mu^{\rm{pd}} \Big|\Big|^2 
+ \sigma W(\Upsilon_k) 
+ \rho \delta_k
 \\
\textrm{s.t.} \,\, & \bar{D}(\bar{q}_k) \ddot{\bar{q}}_k + \bar{H}(\bar{q}_k, \dot{\bar{q}}_k) 
= \bar{B} u_{s, k} + \bar{J}_h^T(\bar{q}_k) \bar{\lambda}_{h, k} + \mathcal{F}^{\rm{avg}}_{k-1}
\\
& L_F\bar{V}^s_\varepsilon(\mathcal{X}_k) + L_G\bar{V}^s_\varepsilon(\mathcal{X}_k) \big(\dot{J}_{y, k} \dot{\bar{q}}_k + J_{y, k} \ddot{\bar{q}}_k \big) \leq - \frac{\gamma}{\varepsilon} \bar{V}^s_\varepsilon(\mathcal{X}_k) + \delta_k.
\\
& -u_{\rm{max}} \leq u_s \leq u_{\rm{max}} 
\end{aligned}
$}
\end{equation}
Although we use the residual dynamics estimate from the previous time step to model the dynamics at the current time step, when run in a controller at a high enough frequency this method should capture the residual dynamics well enough.

\section{Human-Prosthesis Simulation} \label{sec:Simulation}
To demonstrate this ID-CLF-QP+$\mathcal{F}^{\rm{est}}$ we first apply it to a prosthesis model in simulation while the human portion of the system is controlled by a method unknown to the prosthesis. The accuracy of the force estimation is also tested.

\begin{figure} [t] \label{fig:model} 
\centering
\includegraphics[width=1\columnwidth]{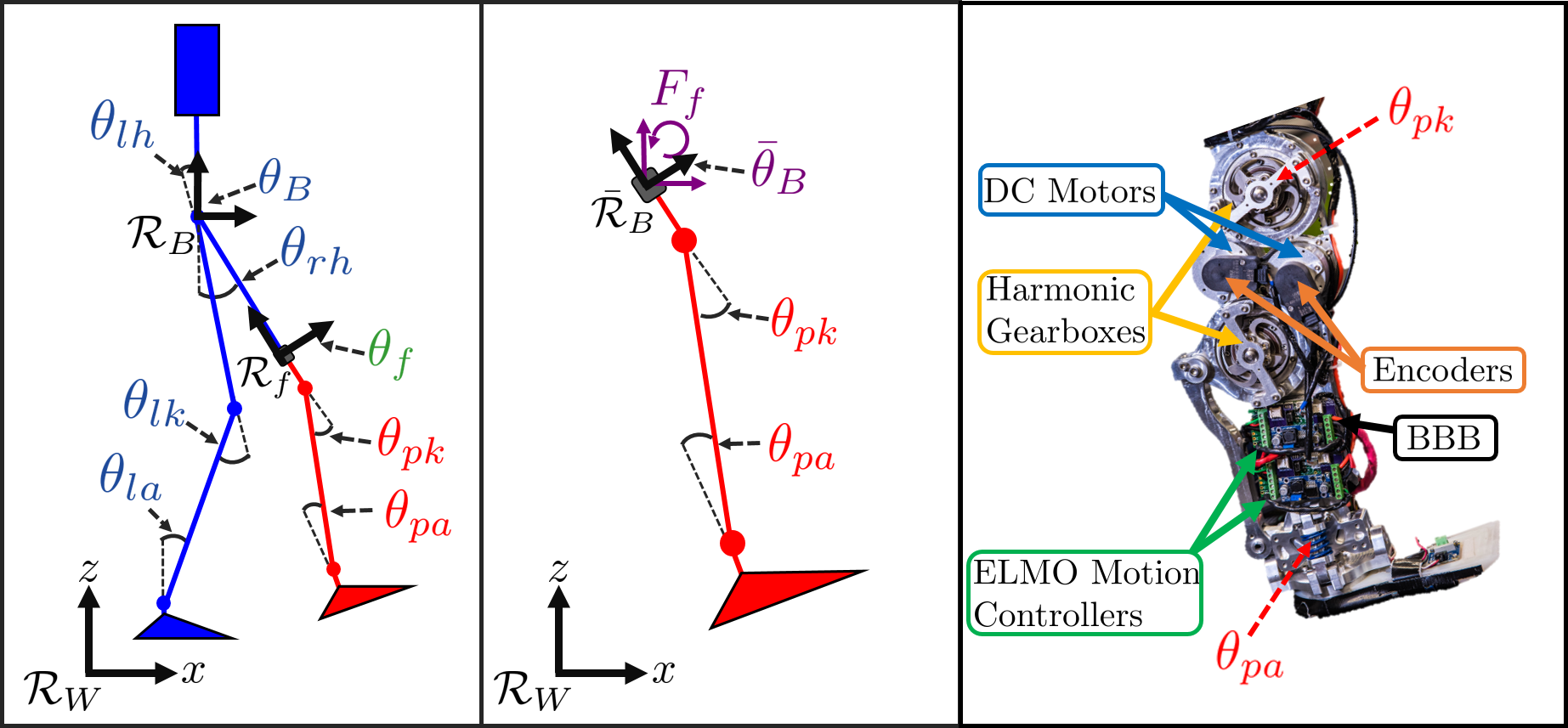}
\vspace{-0.7cm}
{\caption{(Left) Human-prosthesis model with generalized coordinates. (Middle) Prosthesis separable subsystem model with generalized coordinates. (Right) AMPRO3 powered prosthesis platform with components and coordinates labeled.}}
\vspace{-0.7cm}
\end{figure}

\newsec{Amputee-Prosthesis Model.}
We construct an amputee-prosthesis model as a planar bipedal robot comprised of 8 links: torso, 2 human thighs, prosthesis partial thigh, a human and prosthesis calf, and a human and prosthesis foot. The interface between the human right thigh and prosthesis partial thigh is modeled as a 3 DOF fixed joint, as described in Section \ref{sec:Background}, giving $\eta = 12$. The subsystem coordinates of the prosthesis are knee $\theta_{pk}$ and ankle pitch $\theta_{pa}$, $q_s = (\theta_{pk}, \theta_{pa})^T$, giving $\eta_s = 2$. The rest of the system coordinates are the floating base coordinates $\theta_B$, and the pitch of the human's left hip $\theta_{lh}$, left knee $\theta_{lk}$, and right hip $\theta_{rh}$: $q_r = (\theta_B^T, \theta_{lh}, \theta_{lk}, \theta_{rh})^T$. See Fig. 2. All the pitch joints are actuated, making $m_r = 4$ and $m_s = 2$.

The human parameters are obtained with a subject's height and weight and the parameters in \cite{HumanParam, HumanInertia}. The prosthesis parameters are based off of the prosthesis platform AMPRO3 \cite{zhao2017preliminary} used in this study. We use this prosthesis model to obtain the dynamics of \eqref{eq:robotSubsystem} for use in \eqref{eq:ID-CLF-QP+Fest} on the prosthesis platform, but omit the ankles in trajectory generation and simulation because it is more comfortable for the human user to have the prosthesis ankle have varying set point PD control instead of following a trajectory.

\newsec{Hybrid Systems and Human-Like Gait Generation.}
To account for both the continuous and discrete dynamics in human walking, we model it as a hybrid system \cite{ames2014human}. Because the human-prosthesis system is asymmetric, we consider two continuous domains, $\Domain_{\rm{ps}}$ for prosthesis stance and $\Domain_{\rm{pns}}$ for prosthesis non-stance. These domain indices are denoted as $v \in \{{\rm{ps}}, {\rm{pns}}\}$. 
Each domain has a holonomic constraint on the respective stance foot.
The domains are connected by events in a directed graph, specifically the event when the non-stance foot strikes the ground. The impact dynamics for these transitions are explained in \cite{ModelsGrizzle}.

To find a human-like walking trajectory for the model, human walking motion capture data is taken and \Bezier polynomials are fit to the joint trajectories. A state-based phase variable, going from 0 to 1, modulates the trajectories \cite{westervelt2018feedback}. We run an optimization to minimize the difference between the outputs (the joints) while satisfying the dynamics \eqref{eq:robotDynamics}, feasibility constraints, and a hybrid zero dynamics condition \cite{westervelt2018feedback} such that the outputs are invariant through impact. The optimization solution gives parameters to define the desired trajectories $y^d_s(\tau(x_s), \alpha)$ and the outputs to simulate the human side.
See \cite{gehlhar2020data} for details. Fig. 3 shows the resulting trajectories match the human data well. 
By finding a prosthesis knee trajectory similar to a human's knee trajectory and is provably stable when the rest of the system is following the human-like trajectories, we assume the human can still stabilize itself with the prosthesis. Hence the condition required for our main theoretical idea is satisfied.

\begin{figure} [t] \label{fig:human_output} 
\centering
\includegraphics[width=1\columnwidth]{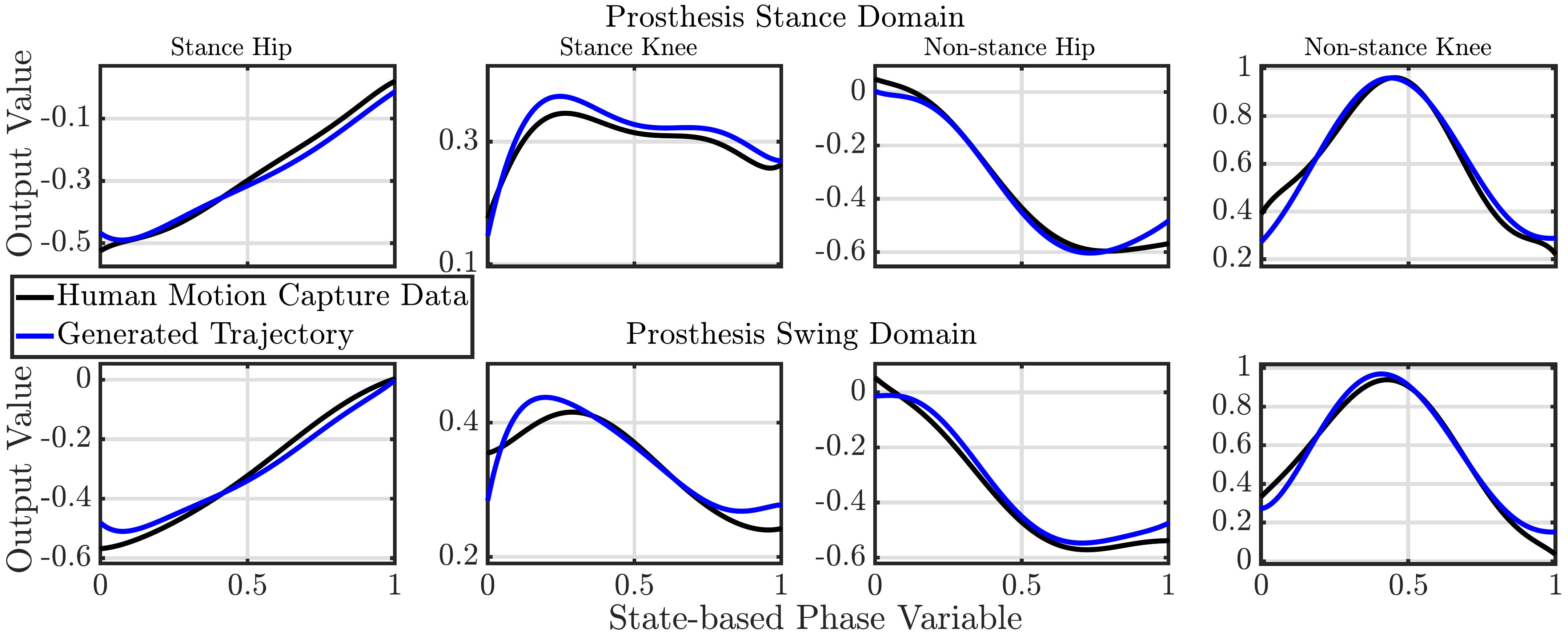}
\vspace{-0.75cm}
{\caption{Joint outputs from optimization (blue) align closely with human motion capture data (black) showing the trajectories we use to test the human-prosthesis model in simulation and implement on the prosthesis device are human-like.}}
\vspace{-0.75cm}
\end{figure}

\newsec{Simulation Results.}
We restrict our attention to implementing the proposed controller in the stance domain $\Domain_{\rm{ps}}$ where the interaction force is the largest and the prosthesis' stability is critical as it supports the human. In practice we calculate the base coordinates $\bar{q}_B$, base velocities $\dot{\bar{q}}_B$, and phase variable $\tau(x_s)$ with inverse kinematics using the knee and ankle data and assuming the foot is flat on the ground. The swing domain $\Domain_{\rm{pns}}$ requires an IMU to provide information about this domain's main unknown, the base coordinates. This remains for future work.

We prescribe a feedback linearizing control law to the human side to closely track the human-like trajectories in simulation. Variations of the ID-CLF-QP controller are implemented on the prosthesis in stance and a feedback linearizing control law in swing to enforce the output \eqref{eq:subOutputs}, where $y^a_s(x_s) = \theta_{pk}$. 
The ID-CLF-QP+$F_f$ is implemented with the exact interaction force $F_f$ calculated with \eqref{eq:force}, since $F_f \in \lambda_h$, based on a feedback linearizing control law $u$. The ID-CLF-QP+$\mathcal{F}^{\rm{est}}$ used the force estimator \eqref{eq:forceavg} with $N = 1$ since averaging is unnecessary in simulation.
Finally the ID-CLF-QP was used without any interaction force information. 

The resultant control inputs are shown in Fig. 4a and tracking results in Fig. 5.
The ID-CLF-QP+$F_f$ and ID-CLF-QP+$\mathcal{F}^{\rm{est}}$ achieved practically exact tracking results and had very similar control inputs.
This suggests the force estimator estimates the force well enough to give similar performance as when using the exact force.
The ID-CLF-QP with no consideration of the interaction force outputs a very different control input and had terrible tracking, indicating the significance of accounting for the force.
To compare the force estimate with the actual computed force, the summation of the constraint wrenches and interaction force projected into joint space is taken since the constraint wrench calculation for the subsystem controller \eqref{eq:ID-CLF-QP+Fest} is coupled with the interaction force estimate and hence they cannot be individually compared with the constraint forces and interaction force calculated with the full-order dynamics \eqref{eq:robotDynamics}. Fig. 4b compares the actual force components calculated by $(\bar{J}_h(\bar{q}) \bar{J}_f(\bar{q}) ) \lambda_h$ to the estimated force components $\bar{J}_f(\bar{q}) ) \bar{\lambda}_h + \mathcal{F}$, showing the force estimation works with high accuracy.

\begin{figure} [t] \label{fig:sim} 
\centering
\includegraphics[width=\columnwidth]{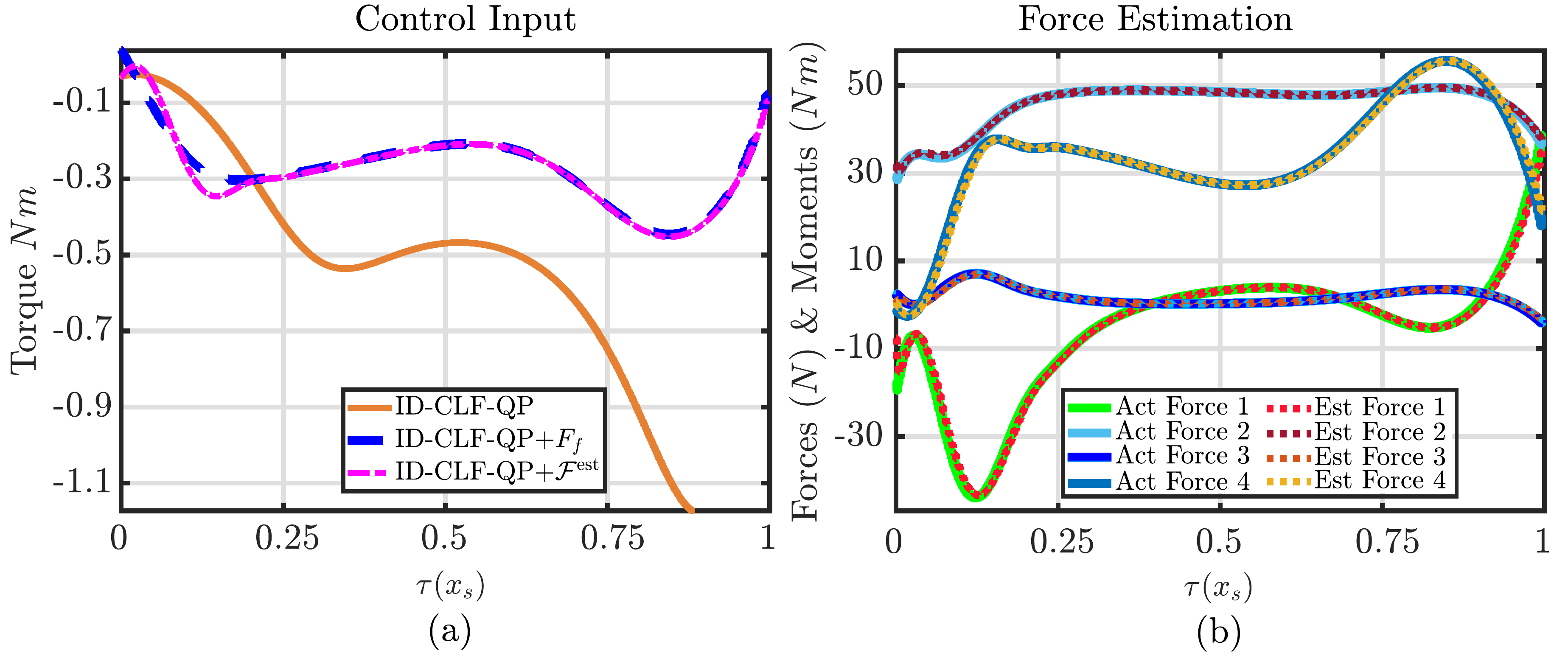}
\vspace{-0.8cm}
{\caption{(a) Prosthesis stance control input for the knee from 3 simulations of the human-prosthesis model walking with variations of the ID-CLF-QP applied to the prosthesis. (b) The summation of the constraint forces and interaction forces projected into joint space. }}
\vspace{-0.8cm}
\end{figure}

\section{Human-Prosthesis Experimentation} \label{sec:Experimentation}
The platform used to demonstrate the model-based control method is described in this section followed by experimental results of the proposed controller.
The results verify this controller meets our formal condition for exponential stability and it outperforms the less model-dependent controllers.

\newsec{Prosthesis Platform AMPRO3.}
The custom-built powered prosthesis AMPRO3 used in this work is described in \cite{zhao2017preliminary}, and briefly described here. The device has an iWalk adapter such that an able-bodied human can test the device. A different adapter can be used to connect this device directly into an amputee's socket. The mechanical design consists of a knee and ankle pitch joint which are both controlled with their own DC brushless motors (MOOG BN23), with about 1 Nm peak torque. Each motor is connected to a harmonic gearbox through a timing belt. Both the timing belt and harmonic gearbox contribute to the mechanical reduction for each joint: 120 for the knee and 175 for the ankle.

Each motor is controlled by an ELMO motion controller (Gold Solo Whistle) which receives position and velocity feedback from incremental encoders and receives input from the microprocessor. A Beaglebone Black (BBB) microprocessor runnning at 200 Hz handles all of the computations on board, taking input from the motion controllers and outputting a commanded torque to the motion controllers. The controller algorithms are coded in C++ packages and run with ROS. The whole prosthesis system is powered by a 9-cell 4400 mAh Li-Po battery (Thunder Power RC). The components described here can be seen in Fig. 2.

\newsec{Hardware Results.}
The ID-CLF-QP+$\mathcal{F}^{\rm{est}}$ was implemented on the prosthesis platform in stance (with $N = 10$ in \eqref{eq:forceavg}) and superior trajectory convergence and tracking were achieved compared to a model-independent PD controller and the ID-CLF-QP controller without consideration for the force. An able-bodied human tested the device in walking for over 20 consecutive steps with each controller. The ankle had a PD controller with varying set point. 
A PD controller was applied to the knee in swing, but did not perfectly converge to the trajectory. Hence the output starts off the trajectory in stance, explaining the jump present in the desired trajectory in Fig. 5. However, the ID-CLF-QP+$\mathcal{F}^{\rm{est}}$ recovers from this disturbance and converges to the trajectory, demonstrating the advantage of the exponential convergence of a model-based RES-CLF. Fig. 5 also shows the significant tracking improvement exhibited by the ID-CLF-QP+$\mathcal{F}^{\rm{est}}$ in stance compared to the other controllers. The rapid convergence and superior tracking are two important results of this work. 

\newsec{Main Result.} The primary result of this work is implementing a model-dependent controller on a prosthesis with formal guarantees of stability. Fig. 6 shows this result where the CLF derivative is plotted with its stability bound, indicating the prosthesis satisfies this formal guarantee of stability. (The slight breaking of the bound is due to the relaxation term in the CLF-QP). When the CLF condition is well below its bound, the control input, shown in the bottom of Fig. 6, has a small magnitude because the controller is letting the natural dynamics of the system bring it to its desired trajectory. This effect is especially significant considering the prosthesis starts off the trajectory at the beginning of the stance phase and this precisely demonstrates the advantage of model-dependent control over model-independent control. A controller without model information would respond to the large error with a large torque which would require more energy and the sudden movement could cause discomfort to the user. This model-dependent controller, on the other hand, allows the natural dynamics of the system to bring the prosthesis to its desired trajectory without using more energy and yielding a less aggressive movement for the user. 

(Note: Due to COVID-19 restrictions, the results of this study are restricted to one subject. Future work will demonstrate the control method on more subjects.)

\begin{figure} [t] \label{fig:outputs} 
\centering
\includegraphics[width=\columnwidth]{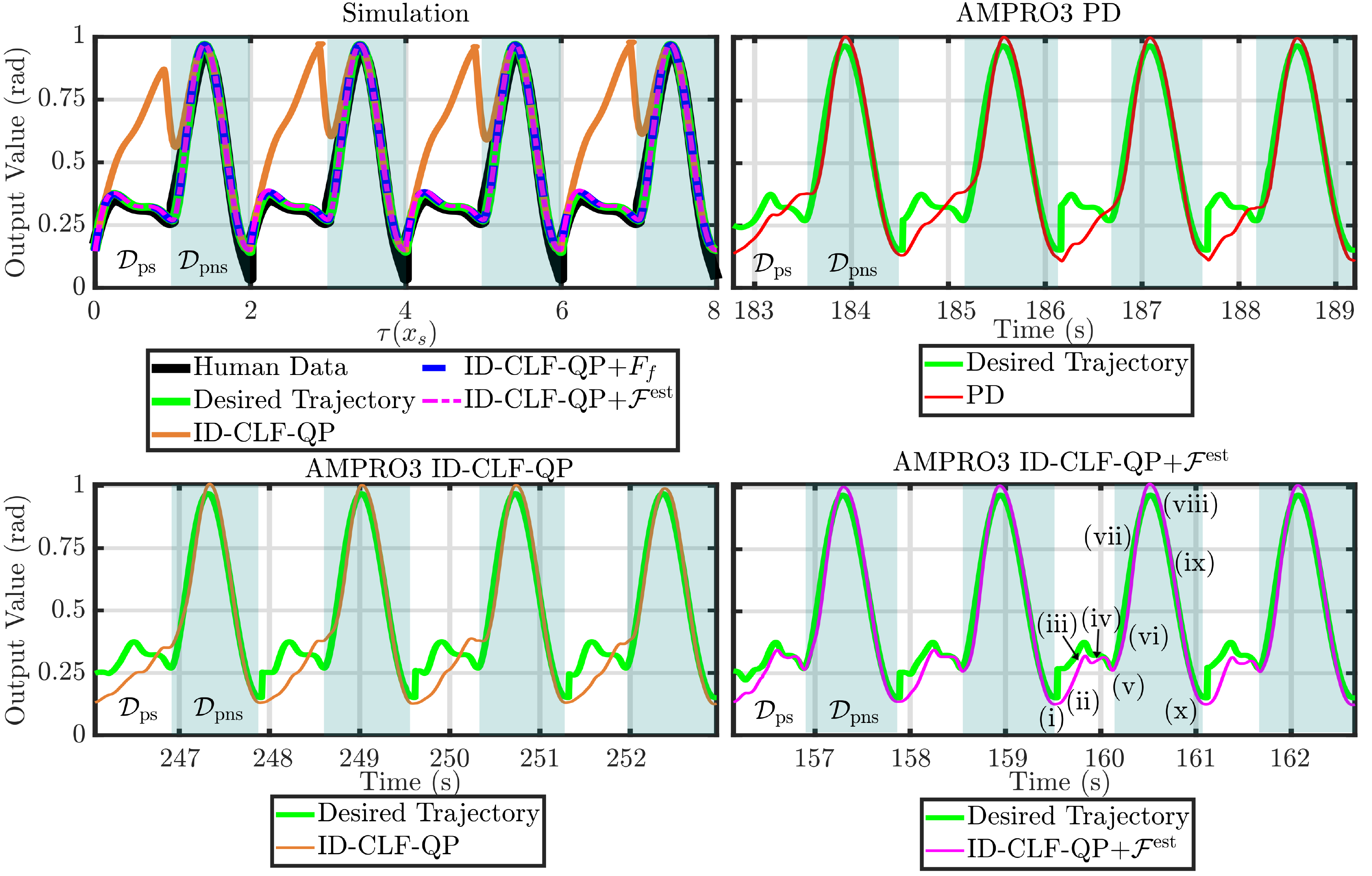}
\vspace{-0.8cm}
{\caption{(Top left) Output tracking from 3 simulations with variations of the ID-CLF-QP on the prosthesis in stance plotted with the desired trajectory and the human data with respect to the phase variable. Experiment output tracking with the PD controller (top right), ID-CLF-QP (bottom left), and ID-CLF-QP+$\mathcal{F}^{\rm{est}}$ (bottom right) applied in stance plotted with the desired trajectory in time. $\Domain_{\rm{ps}}$ white, $\Domain_{\rm{pns}}$ shaded. Numbers in bottom right plot indicate phase of gait corresponding to gait tiles in Fig. 1}}
\vspace{-0.3cm}
\end{figure}

\begin{figure} [t] \label{fig:ampro_clf} 
\centering
\includegraphics[width=\columnwidth]{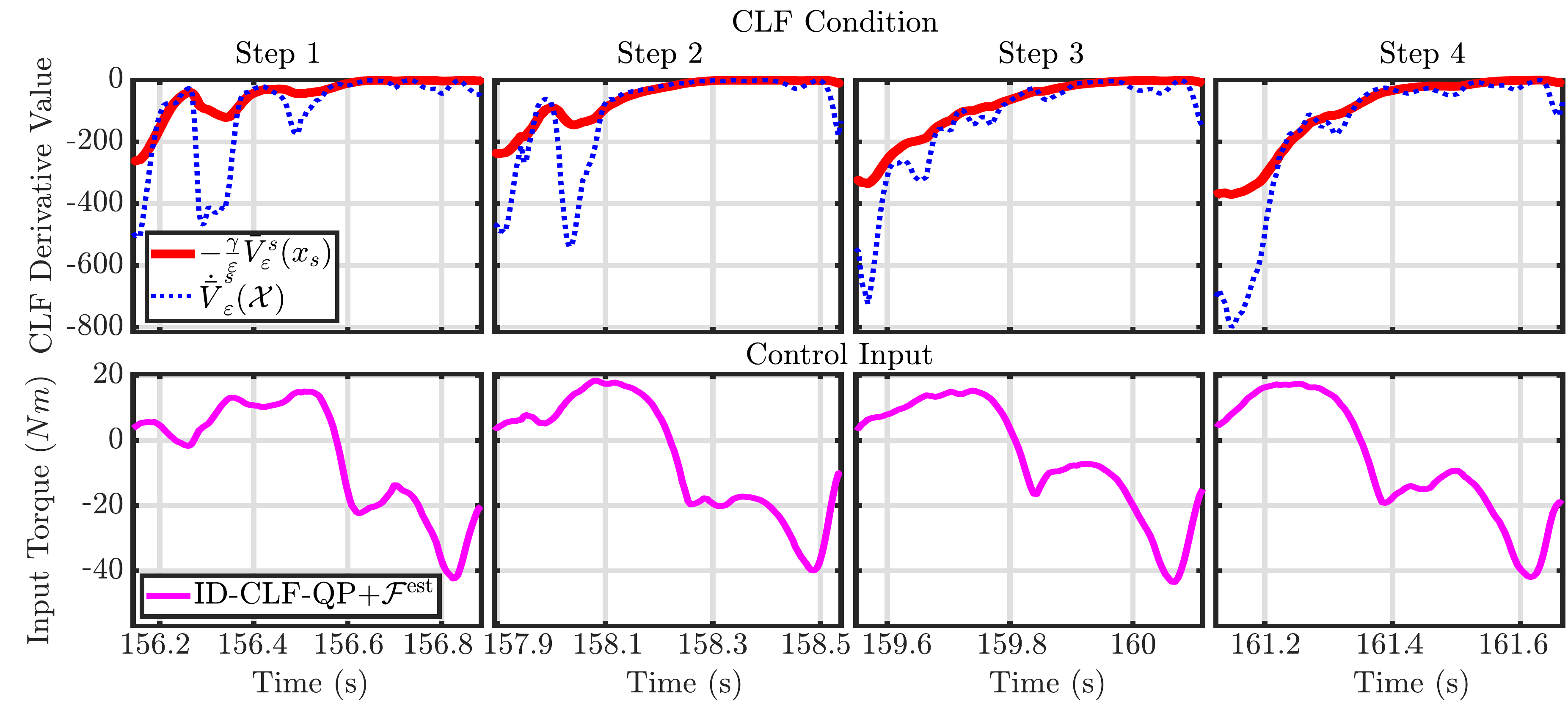}
\vspace{-0.7cm}
{\caption{ Results of four phases of stance from experiment. 
(Top) The RES-CLF derivative (blue) plotted against its bound (red). (Bottom) The prosthesis knee control input.}} 
\vspace{-0.7cm}
\end{figure}

\section{Conclusion and Future Work} \label{sec:Conclusion}
In this work, the novel methodology of developing RES-CLFs for separable systems \cite{gehlhar2019control, gehlhar2021separable} is realized on a prosthesis platform, demonstrating the first experimental realization of a model-dependent prosthesis controller that accounts for interaction forces. As such, this is the first instance of realizing prosthesis control with formal guarantees of stability for the full-order hybrid system with zero dynamics.  These guarantees with consideration for the interaction forces ensure safety of the user and a responsiveness to the real-time dynamics are novel relative to existing prosthesis control methods.
Being able to implement model-dependent controllers on a prosthesis platform opens the door to applying various nonlinear control techniques to prostheses and other robotic subsystems, thereby improving performance.

Future work will apply this control method in the swing phase by incorporating an IMU into the prosthesis platform. Ways to improve the accuracy of the force estimation method will also be investigated. Episodic learning could adapt our force estimate to systematically reduce uncertainty while maintaining stability, similar to the work of \cite{taylor2019episodic}. Including a force sensor at the socket could measure the interaction force in real-time and using the estimation method presented in this paper along with the aforementioned learning method could address the issues of measurement noise and uncertainty.

\newpage
\bibliographystyle{IEEEtran}
\balance
\bibliography{bibliography}

\end{document}